\pdfoutput=1 
\documentclass[12pt]{article}
\usepackage{amsmath,amsfonts,amssymb,a4,color,epsfig,graphics}
\usepackage[latin1]{inputenc}
\newcommand{\R}{{\mathbb{R}}}

\newcommand{\C}{{\mathbb{C}}}
\newcommand{\Z}{{\mathbb{Z}}}
\newcommand{\N}{{\mathbb{N}}}

\def\pa{\partial}
\def\ra{\rightarrow}
\def\ga{\alpha}
\def\gb{\beta}

\def\ge{\varepsilon}

\def\gg{\gamma}

\def\gl{\lambda}

\def\gG{\Gamma }

\newtheorem{defi}{Definition}[section]

\newtheorem{lemm}{Lemma}[section]
\newtheorem{prop}{Proposition}[section]
\newtheorem{rem}{Remark}[section]
\newtheorem{coro}{Corollary}[section]
\newtheorem{theo}{Theorem}[section]
\newtheorem{exem}{Example}[section]
\newtheorem{conj}{Conjecture}[section]

\begin{document}

\title{Topological Resonances on Quantum Graphs}
\author{Yves  Colin de Verdi\`ere\footnote{Universit\'e de Grenoble-Alpes,
Institut Fourier,
 Unit{\'e} mixte
 de recherche CNRS-UJF 5582,
 BP 74, 38402-Saint Martin d'H\`eres Cedex (France);
{\color{blue} {\tt yves.colin-de-verdiere@univ-grenoble-alpes.fr}}}\\
\& Fran\c coise Truc}

\maketitle

\section*{Introduction}
We will consider metric  graphs $G$ which consist of a finite graph $\Gamma $ with some leads attached to some
vertices. To this metric graph is associated a Laplacian using the Kirchhoff conditions.
Resonances on such Quantum graphs are introduced in the book \cite{BK13} and studied in several papers (see \cite{EL10,DP11,LZ16}). 
In the paper \cite{DP11}, the following result is proved:
\begin{theo}\label{theo:davies}
All resonances $k_j=\sigma_j +i\tau_j,~j\in\N, $ ly in a band
$-M\leq \tau_j\leq 0$ and they have the large $K$ asymptotic
\[ \# \{j~|~ | \sigma_j| \leq K \}= \frac{2 V}{\pi} K + O(1)~,\]
with $0 <  V \leq |L| $ where $|L|$ is the total length of the finite graph $\Gamma$.
\end{theo}

In this paper, we will be interested in resonances close to the real axis which in physics
are the most important. They are linked to compactly supported eigenfunctions as anticipated in \cite{EL10}.
The goal of this paper  is to describe some asymptotic properties of  these resonances, called
``topological resonances'' in the paper
\cite{GSS13}. See also the paper \cite{LZ16} for the explicit calculations of the related ``Fermi golden rule''.
We show that there is a dichotomy between graphs which can have eigenfunctions with compact support for some particular metrics 
 and the other ones
which are some specific  trees, namely those  with at most one vertex of degree one.
In the first case, there are many resonances close to the real axis and we are able to say something on their asymptotics, while 
in the second one, there is a gap which is an invariant associated to the graph.

We will follow the notations of the paper \cite{CdV15} of the first author.

{\it Acknowledgements: the first author thanks Uzy Smilansky for motivating him to  study the topological resonances while visiting our
 Institut. This work has been partially supported by the LabEx PERSYVAL-Lab (ANR--11-LABX-0025). }

\section{Main results and conjectures}
We consider a finite  graph $\Gamma =(V,E)$  and fix a subset $V_0 \subset V$.
To each vertex $w$ of $V_0$, we attach a non zero number $n_w $ of   infinite half-lines, called  the  ``leads'' in \cite{GSS13}.
The total number of leads is denoted by $N$  with $N=\sum_{w\in V_0} n_w$.

Allowing loops and multiple edges,
we can (and will) always assume that 
\[ \forall v\in V,~{\rm degree}_G (v) \ne 2 ~.\]
The metric graph denoted by $G $  is the union 
of $\Gamma $ with some lengths $l_e>0  $ for $e\in E$,  and the attached infinite half-lines.
 Let us denote by $\vec{l}$ 
the vector in  $\R^n$ with coordinates $l_e,~e\in E$.
To this set of data, we associate, using the usual Kirchhoff conditions at the vertices, a non negative self-adjoint 
 Laplacian $\Delta _G ^{\vec{l}}$ acting
on $L^2(|G|,|dx|_{\vec{l}})$ where $|G|$ is the 1D singular manifold associated to $G$
and $|dx|_{\vec{l}} $ is the Riemannian measure. Let us specify  that, at any  vertex of degree $1$, we impose Neumann boundary conditions.
This Laplacian  has a discrete sequence  of non negative eigenvalues (possibly empty) and a continuous spectrum $[0,+\infty[$
of multiplicity $N $.
The Schwartz kernel of the resolvent $\left(k^2 -\Delta_G ^{\vec{l}} \right)^{-1}$
 defined for $\Im k >0 $
extends to the lower half plane in a meromorphic way.
The poles of this extension in $\Im k\leq 0$ are called the {\it resonances}.
We denote by ${\rm  Res}_G^{\vec{l}}$ the set of resonances of  $\Delta _G ^{\vec{l}}$.
Our goal is to study ${\rm  Res}_{G}^{\vec{l}}$, mainly in the case where the lengths $(l_e)_{e\in E}$ are independent
 over the integers (we will say that $\vec{l}$ is ``irrational'').

There are two mutually disjoint families of graphs $G $:
\begin{itemize}
\item {\it Type I:} the trees with at most one vertex of degree $1$
\item {\it Type II:} all other graphs.
\end{itemize}
The {\it Type I} graphs are the graphs $G$ for which, for any choice of $\vec{l}$, the operator
$\Delta_G ^{\vec{l}}$ has   no $L^2$ eigenfunctions or equivalently $\Delta_\Gamma  ^{\vec{l}}$ has no eigenfunctions vanishing on $V_0$.

The main results of this note are
\begin{theo}\label{theo:trees}
 If the graph $G$ is a tree with at most one vertex of degree $1$ ($G$ is of type I),
there exists a minimal finite number 
$h(G)>0$ so that, for any choice of the lengths $l_e>0$ with $|L|:=\sum l_e$, we have
\[  {\rm  Res}_G^{\vec{l}} \subset \{\sigma +i\tau ~|~ \tau |L| \leq - h(G) \} ~.\]
\end{theo}
The optimal constant $h(G)$ is 
an interesting graph parameter. It follows from the next results that $h(G)=0$ for type II graphs.
Moreover we have that 
\begin{theo} If $G'$ is obtained from $G$ by contracting some edges of $\Gamma $ while keeping the leads in a natural way, then 
$h(G')\geq h(G)$.
\end{theo}
It would be interesting to say more on this graph parameter.

In order to state the second main result, we need the 
\begin{defi} \label{defi:wg} For a graph $G$ as before, 
$W_G$  is the sub-set of the torus $T^E:=(e^{il_e})_{e\in E}$ so that $\Delta _\Gamma^{\vec{l}}$ admits  $1$ as an 
  eigenvalue with a non zero   eigenfunction vanishing on $V_0$; 
 we define also $W_G^o$ as  the subset of $W_G$ where the  eigenfunction of $\Delta _\Gamma^{\vec{l}}$ vanishing on $V_0$  is 
  unique, up to scaling.
\end{defi}
Note that $W_G$ depends only on the choice of $V_0$.
If $G$ is of type II, the sets $W_G$ and $W_G^o$   are  non-empty semi-algebraic sets (see \cite{CdV15}
and Appendices \ref{app:nd} and \ref{app:alg}). It follows from Theorem \ref{theo:tangent} that $W_G^o$ is a smooth 
(non closed in general) submanifold of $T^E$ 
which is an union of connected components (called the strata) of various dimensions. The maximal dimension of these strata is called 
the dimension of $W_G^o$.
We will need the
\begin{defi}
The number $d(G)$ is defined by 
\[ d(G):=\# E -1 - \dim W_G^o~.\]
\end{defi}
We will be interested in the following quantity:
\begin{defi} For $\ge \geq 0 $, let us define   $N_{G,\vec{l}}(\ge)$ as
follows:
\[~  N_{G,\vec{l}}(\ge):=\liminf_{K \ra +\infty} \frac{1}{K}
\# \{ \sigma_j+i\tau_j \in {\rm  Res}_{G}^{\vec{l}}~|~ 0\leq \sigma_j \leq K,~-\ge \leq \tau_j \leq 0 \}~. \]
\end{defi}
Finally let us define
a combinatorial invariant $g(G)$ of the graph $G$:
\begin{defi} If $G$ is of type I, $g(G)=+\infty$, otherwise
$g(G)$ is the smallest number so that there exists either a simple cycle in $G$ with $g(G)$ vertices or
a path in $G$ joining two vertices of degree 1 with   $g(G)+1$ vertices (and $g(G)$ edges).
\end{defi}
The number $g(G)$ can be smaller than the girth of $G$, for example if $G$ is a tree of type II.

We have the
\begin{theo} \label{theo:type2}
\begin{enumerate}
\item
If the graph $G$ is of type II  and
  $\vec{l}$ is irrational, 
there exist $\ge_0 >0$ and $C>0$, depending on $\vec{l}$, so that, for $0<\ge \leq \ge_0$, we have
\[ N_{G,\vec{l}}(\ge)\geq  C\ge ^{d(G)/2} ~.\]
\item
Moreover, we have
\begin{itemize}
\item If there exists $a\in V_0 $ so that there is no loops at the vertex $a$, we have $d(G)\geq 1$
\item  $d(G) \leq  g(G)-1 $
\item If  $\gamma  $ is a simple cycle of $\Gamma $, then
\begin{itemize} \item
 either 
$d(G) \leq \# V_0 $ for all  $V_0 \subset V(\gamma )$
\item or there exists a vertex $a$ of $\gamma $ so that $d(G) \leq  \# V_0 $
if 
 $V_0 \subset V(\gamma )$  and  $a\notin V_0$.
\end{itemize}
\item If $\Gamma $ is 2-connected,
\begin{itemize}
\item either $d(G)=1$ for all sets $V_0$ with $\# V_0=1$
\item or  there exists  a unique  vertex ${a}$ of $\Gamma $ so that
$d(G)= 1 $ for all sets $V_0$ with $\# V_0=1$ and  $V_0 \ne \{ a \} $.
\end{itemize}
\item If $ V_0=V$, then $d(G)=g(G)-1$.
\end{itemize}
\end{enumerate}
\end{theo}

We are not able to derive an upper bound in general, but we conjecture, following \cite{GSS13},
the following estimates: 
\begin{conj}\label{conj:main} As $\ge \ra 0^+$, there exists $C>0$ so that 
\[   N_{G,\vec{l}}(\ge) \sim  C\ge ^{  d(G)/2}~,\]
and 
\[ d(G)=\min (g(G)-1, \# V_0 )~.\]
\end{conj}

\begin{rem} {\bf How is $N_{G,\vec{l}}(\ge)$  related to the Gnutzmann-Schanz-Smilansky paper \cite{GSS13}? }
Let us choose the resonant state $u_j$ associated to $k_j=\sigma_j+i\tau_j$ so that 
$\sum_{m=1}^N  |t_m|^2=1$,  then we look at its energy in $\Gamma $ defined
by
 \[ {\cal E}(u):=\int _{|\Gamma |} |u|^2 |dx|_{\vec{l}}~.\]
Proposition  \ref{prop:norm} gives that $ {\cal E}(u)=1/(2|\tau|)$.
The previous authors look at
the asymptotic behaviour, as $\ga \ra \infty $,  of
\[ P_{G,\vec{l}}(\ga ):=\lim_{K \ra \infty} \frac{1}{K} \# \{ k_j=\sigma_j +i\tau_j| 0\leq \sigma_j \leq K, ~ {\cal E}(u_j)\geq  \ga \}~,\]
which is the same as $N_{G,\vec{l}}(1/2\ga)$ at which we are looking in the present paper.

\end{rem}

\section{Finding resonances}
We will use the following notations: if $z$ is a vector in $\C^n$, $(z)$ will be the diagonal matrix with entries the coordinates of $z$
and $(z)_2$ will denote the diagonal matrix of size $2n$ 
\[ (z)_2= \left( \begin{matrix} (z) &0 \\ 0& (z) \end{matrix} \right)~.\]

 We choose  an orientation of each edge of $\Gamma $ and parametrize
the edge $e$ by a real parameter $x_e$ with $0\leq x_e \leq l_e $ according to the orientation. The leads
are parametrized by $x_m$~,~$ m=1,\cdots ,N,$ with  $0\leq x_m <\infty$.
We will denote by $|dx|_{\vec{l}}= \sum _{e\in E} |dx_e|+ \sum_{m=1}^N |dx_m| $ the Riemannian measure on $|G|$ where
$|G|$ is the 1D singular topological space associated to $G$.
\begin{defi} A complex number $k=\sigma +i\tau $ with $\tau \leq 0$ is a ``resonance'' of $ \Delta_G ^{\vec{l}}$ if and  only if 
there exists a non zero ``resonant state'' $u$ which satisfies $''(\Delta_G ^{\vec{l}}-k^2 )u=0'' $
and $\forall m=1, \cdots ,N ,~\exists t_m \in \C, ~ u(x_m)=t_m {\rm exp}(ikx_m)$.\end{defi}

Equivalent  definitions of resonances for Quantum graphs are given  in \cite{EL07}.

Following \cite{BK13} sec. 5.4., we can describe all solutions of $(\Delta_G^{\vec{l}} -k^2)  u=0$
in the following way: writing $u(x_e)=a_e {\rm exp}(ikx_e) + b_e {\rm exp}(-ikx_e)$
and $u(x_m)=t_m^{\rm out} {\rm exp}(ikx_m )+t_m^{\rm in} {\rm exp}(-ikx_m ) $, and denoting
 by $e^{ik\vec{l}}$ the point in the $\C^E$  of coordinates $e^{ikl_e}~, e\in E$, 
the Kirchhoff conditions at the vertices express as
\begin{equation} \label{equ:S} \left( \begin{matrix} t^{\rm out}\\ a \\b \end{matrix} \right)=
\left( \begin{matrix} R & T_o\\ T_i & U\left(e^{ik\vec{l}}\right)_2 \end{matrix} \right)
\left( \begin{matrix} t^{\rm in}\\ a \\b \end{matrix} \right)~. \end{equation}

The $2n+N$ square matrix 
\[ S = \left( \begin{matrix} R & T_o\\ T_i & U\left(e^{ik\vec{l}}\right)_2 \end{matrix} \right)\]
is unitary for real $k$'s.
The resonances are  the value of $k$ for which there exists a non trivial solution of
Equation (\ref{equ:S})  with $t^{\rm in}=0$ and are hence  given by the equation 
\[ {\cal R}_G( {\rm exp}(ikl_1), \cdots,  {\rm exp}(ikl_n))=0\]
where
\[  {\cal R}_G (z)= {\rm det} ({\rm Id} - U (z)_2)~.\]
The associated resonant state is given by
\[  U\left(e^{ik\vec{l}}\right)_2\left( \begin{matrix}  a \\b \end{matrix} \right)=\left( \begin{matrix}  a \\b \end{matrix} \right)~,\]
\[ t^{\rm in}=0,~t^{\rm out}=T_o \left( \begin{matrix}  a \\b \end{matrix} \right)~.\]

We will need the
\begin{prop}\label{prop:norm}
If $u$ is a resonant state for the resonance $k=\sigma +i\tau $ of $ \Delta_G ^{\vec{l}}$
with $\sigma \ne 0$,
we have
\[ -2\tau \int _{|\Gamma |} |u|^2 |dx|_{\vec{l}} =  \sum _{m=1}^N |t_m|^2 ~.\]
\end{prop}
{\it Proof.--}
Let us evaluate the integral
\[ I= \int_{|\Gamma |} \left( u \Delta \bar{u}-\bar{u} \Delta u \right) |dx|_{\vec{l}} \]
in two ways:
first, using the differential equation $ \Delta u =k^2 u $,
we have 
\[ I= -4i\sigma \tau \int_{|\Gamma |}|u|^2 |dx|_{\vec{l}}~,\]
whereas using the Green-Riemann formula, we get:
\[ I= 2i\sigma \sum _{k=1}^N|t_m|^2 ~ . \]
\hfill  $\square $
\begin{coro}
The   resonant states with $\tau =0$ are   eigenstates of $\Delta _G ^{\vec{l}}$
with support in $|\Gamma |$.
And conversely, each such eigenfunction of $\Delta _G ^{\vec{l}}$
is a resonant state of  $\Delta _G^{\vec{l}}$.
\end{coro}
Proposition \ref{prop:norm}  implies that all $t_m$'s vanish, hence the conclusions.

\section{Type I:  trees with at most one vertex of degree one}\label{sec:tree}
Let us prove Theorem \ref{theo:trees}.
We start with the following Lemma:
\begin{lemm}\label{noeigen} If $G$ is a tree with at most one vertex of degree $1$ and ${\vec{l}} $ is given, 
then $\Delta _G^{\vec{l}}$ has no non vanishing $L^2$ eigenfunction.
\end{lemm}
{\it Proof.--} Let us assume that the eigenvalue is $k^2>0$.
It is clear that the corresponding eigenfunction $u$ has to vanish on the leads because $u$ is of the
form $a_i \cos kx_i +b_i \sin kx_i $ on $l_i$ and is square integrable. 
 Let us assume that $u$ does not vanish identically  along an edge  $e_1=[v_1,v_2]$, then,
either $v_2$ is of degree 1, or there exists an edge $e_2=[v_2,v_3]$ with $v_3 \ne v_1$, so 
that $u$ does not vanish identically on $e_2$.
Iterating, and possibly going in the other direction,  we get that there exists either a cycle, or a path joigning two vertices of degree 1
so that $u$ has a finite number of zeroes  on them. By assumption, $G$ has no cycles, and  all paths starting from a vertex of degree 1
will end into a lead where it has to vanish identically. $ \square $

Let us now give the proof of Theorem \ref{theo:trees}:

{\it Proof.--}
By rescaling the lengths $l_e$, we can assume that $|L|=1$. By contradiction, there
exists a sequence of vectors $\vec{l}_n$ so that  $u_n$ is  a resonant state of $\Delta_n:=\Delta _G^{\vec{l}_n}$ of resonance
$k_n=\sigma_n +i\tau_n $,
with $\int_{|\Gamma|}|u_n|^2 |dx|_n=1 $, and $\tau_n \ra 0$.
We have the equations
\[ (U\left( z_n \right)_2-{\rm Id}) \left( \begin{matrix}  a_n \\b_n \end{matrix} \right)=0,~
t_n^{\rm out}=T_o \left( \begin{matrix}  a_n \\b_n \end{matrix} \right)~,\]
with
\[ (z_n)_e=e^{ik_n (l_n)_e}~.\]

From Proposition \ref{prop:norm}  we already know that $\vec{t_n^{\rm out}}\ra 0$. We can extract converging sub-sequences of the vectors 
${\rm exp}(i\sigma_n l_e^n)$ going to ${\rm exp}(i l_e^\infty)$ with $2\pi \leq l_e^\infty <4\pi$.
It is not possible that 
the vectors  $({a}_n,{b}_n)$ tend  to $0$, because it would  imply that the $L^2$ norms of $u_n$ on $|\Gamma |$ tend to $0$.
On the other hand, if we denote by $M_n^2 = \max_e  \left(|a_e^n|^2 +|b_e^n|^2 \right)$, the sequence  $({a}_n,{b}_n)/M_n$ 
 converges (up to extraction of a subsequence) to the coefficients of an  eigenfunction with compact support of $\Delta _G^{\vec{l}_\infty}$
with eigenvalue $1$ and  the contradiction follows from Lemma \ref{noeigen}. $ \square $

\begin{exem}We  compute $h(G)$ defined in Theorem \ref{theo:trees} in some simple examples:
\begin{itemize}
\item  $G$ is a star graph where all edges are leads:  there is no resonances.
\item $G$ is a star graph with only one edge of finite length $l$ and $N>1$ leads:
the resonances are 
\[ k_j=\frac{1}{2l}\left( (1+2j)\pi -i \log \frac{N+1}{N-1}\right) ,~j\in\Z ~,\]
so that $h(G)= \frac{1}{2}\log \frac{N+1}{N-1}$.
\item $\Gamma $ is an interval $[v,v']$ of length $l>0$  and $G_{N,N'}$, with $N,N' \geq 2$,  is obtained 
from $\Gamma $ by attaching $N$ leads to $v$ and $N'$ leads to $v'$.
The set of resonances is
\[ \left\{ \frac{1}{2l} \left( 2\pi j -i \log \frac{(N+1)(N'+1)}{(N-1)(N'-1)}\right)~|~ j\in \Z \right\}~\]
and 
\[ h(G)=  \frac{1}{2}\log \frac{(N+1)(N'+1)}{(N-1)(N'-1)}~.\]

\end{itemize} 
\end{exem}

\section{Graphs of Type II}
The goal of this section  is
to describe in a precise way the asymptotic behaviour of the resonances
in the type II  case.

\subsection{Geometric preliminaries}
Let us start introducing some algebraic sets:
\begin{itemize}
%\item $R_G:= {\cal R}_G^{-1} (0)\subset  \C^n$
\item If $\vec{l}$ is given, 
  \[ R_G^{\vec{l}}:=\{ (z_e)_{e\in E} =(e^{i\alpha_e -\tau l_e})_{e\in E} \in {\cal R}_G^{-1}(0),~\alpha _e,~\tau \in \R  \}
\subset T^E \times \R ~.\]
We denote by $Y_G^{\vec{l}}$ the projection of  $R_G^{\vec{l}}$ onto the torus $T^E$:
\[Y_G^{\vec{l}}:=\{ (e^{i\alpha_e })_{e\in E}~|~\exists ~\tau \in\R~ {\rm with~} (e^{i\alpha_e -\tau l_e})_{e\in E}\in R_G^{\vec{l}} \} ~.\] 
\item $Z_\Gamma $ is the determinant manifold of $\Gamma $ defined by
$(e^{il_e})_{e\in E}\in Z_\Gamma $ if and only if $1$ is an eigenvalue of
$\Delta _\Gamma ^{\vec{l}}$.
%\item $W_G\subset  Z_\Gamma $ is the set of points $(e^{il_e})_{e\in E}$ in $Z_\Gamma$ so that
%$\Delta _\Gamma   ^{\vec{l}}$ admits a unique (up to scaling) eigenfunction of eigenvalue 1
%vanishing on $V_0$.
% We remark that $W_G$ depends only on $V_0$ and not of the number of leads attached to each vertex in $V_0$.
\end{itemize}

It follows from Section 4 of \cite{CdV15} that 
\begin{prop}
The graph $G$ is of type II if and only if $W_{G }^o $ is non empty.
\end{prop}
For type I, $W_G$ is empty.
For type II, except for circles,  the first author constructed, in Section 4 of \cite{CdV15}, 
 non zero  eigenstates associated to a non degenerate eigenvalue
 of some
$\Delta_\Gamma^{\vec{l}}$ which vanish on all vertices of $\Gamma $,  except may be two  vertices of degree $1$:
in these cases, $W_{G }^o $ is non empty. 
In the case of the circular graphs (i.e. $|\Gamma |$ is a circle), the eigenspaces of $\Delta _\Gamma $
 are  degenerate, but the resonant states are still non degenerate.

We have the 
\begin{theo}\label{theo:tangent}

If $w_0\in W_{G }^o$, then $ {R_G^{\vec{l}}}$ is smooth near $(w_0,0)$ and, if $u_0$ is
a corresponding eigenfunction with coefficient $(a_e,b_e)_{e\in E}$, then
\[T_{w_0,0} {R_G^{\vec{l}}}= \left\{ \sum_e (|a_e|^2 +|b_e|^2)d\ga_e =0,~d\tau =0 \right\} ~.\]
Similarly, $Y_G^{\vec{l}}$ is smooth near $w_0$ and 
\[T_{w_0} {Y_G^{\vec{l}}}= \left\{ \sum_e (|a_e|^2 +|b_e|^2)d\ga_e =0 \right\} ~.\]

If $u_0$ is also  non degenerate as an eigenfunction of $\Delta _\Gamma^{\vec{l}}$, then 
\[T_{w_0,0} {R_G^{\vec{l}}}= T_{w_0}Z_\Gamma \oplus 0 ~,\]
and 
\[ T_{w_0} {Y_G^{\vec{l}}}= T_{w_0}Z_\Gamma ~.\]
 The set  $R_G^{\vec{l}}$ is near  $w_0$ a graph $\tau =\tau (e^{i\ga_e})$ with $\left(e^{i\ga_e}\right) \in Y_G^{\vec{l}}$.

The differential of $\tau $ vanishes at $w_0$, so that the
Hessian of $\tau $ is well defined on $T_{w_0}{Y_G^{\vec{l}}}$.
\end{theo}

{\it Proof.--} 
Let $w_0=\left(e^{i\ga_e^0}\right)$ be a point in $W_G^o$. The eigenvalue $1$ of $U\left(w_0\right)_2 $ is non degenerate and we choose an eigenvector
$u_0 \in \C^{2E}$ of norm $1$ so that $U\left(w_0\right)_2 u_0=u_0$.
If $w \in \C^E $ is close to $w_0 $, $U(w)_2$ admits an unique eigenvalue $\gl(w)$ close to $1$ which is  non degenerate, and the
associated eigenvector of norm $1$, $u(w)$, is smooth w.r. to $w$.
Let us assume that $w(t)$ depends smoothly of $t$ with $w(0)=w_0$ and compute the derivative
$\dot{\gl } $ of $\gl(w(t)) $ at $t=0$.
Taking the derivative of the equation $U(w(t))_2u(w(t))=\gl(t) u(w(t))$
 w.r. to $t$ at $t=0$, 
 we get with natural notations 
\begin{equation}\label{equ:der}
 U\left(\dot{w}_0\right)_2 u_0 + U\left(w_0\right)_2 \dot{u}_0= \dot{\gl}_0u_0 +\dot{u}_0~.\end{equation}
Taking the scalar product of both sides of Equation (\ref{equ:der})  with $U(w_0)_2 u_0=u_0$, we get, after  simplifications
and using the fact that we can choose $u(t)$ so that  
$< \dot{u}_0 |u_0> =0$, 
\[ \dot{\gl}_0= \langle U\left((w_0) (w_0)^{-1}\dot{w}_0\right)_2 u_0 | U\left(w_0\right)_2 u_0 \rangle
+ \langle U\left(w_0\right)_2 \dot{u}_0| U\left(w_0\right)_2 u_0 \rangle~.\]
Now we can use the identity
\[ \left\langle S  \left( \begin{matrix} 0 \\ v \end{matrix} \right) | S \left( \begin{matrix} 0 \\ u_0 \end{matrix} \right)\right\rangle 
= \langle U\left(w_0\right)_2 v  | U\left(w_0\right)_2 u_0 \rangle~.\]
Because $S$ is unitary at $w_0$, we get
\[ \langle U\left(w_0\right)_2 v  | U\left(w_0\right)_2 u_0 \rangle =\langle v  |u_0 \rangle ~.\]
Using this, we get
\begin{equation}\label{equ:der2} \dot{\gl}_0=\langle \left((w_0)^{-1}\dot{w}_0\right)_2 u_0| u_0 \rangle ~.\end{equation}
Now let us take, with $\vec{l}$ fixed
$w(\ga,\tau)= \left( e^{i\alpha _e -\tau l_e } \right)$, and compute the derivatives of $\gl$ w.r. to $\ga $ and $\tau $ at
$w_0 $; from  Equation (\ref{equ:der2}), 
we get 
\[ \frac{\pa \gl}{\pa \tau}= -\sum l_e m_e \]
and
\[ \frac{\pa \gl}{\pa \ga_e}= i m_e \]
with $m_e= |a_e|^2 + |b_e|^2$.
Now using the fact that the determinant is the product of all eigenvalues, we get that the set ${\rm det }\left(U(w)_2-{\rm Id}\right)=0 $
is also defined by $\gl(w)=1$ with the same non degeneracy properties.
This gives the fact that $R_G^{\vec{l}} $ is near $\left(w_0,0\right)$ a submanifold of codimension 2 of $T^E  \oplus \R _\tau $
whose tangent space is what is given in the Theorem \ref{theo:tangent}. $ \square $

\begin{theo}\label{theo:emb}
For almost all choices of $\vec{l}$, $\Delta _G^{\vec{l}}$ has no $L^2$ eigenfunctions; in particular $N_{G,{\vec{l}}} (\ge =0)=0$.
\end{theo}
This is simply because $W_G^o$ is of codimension at least $1$ in $Z_\Gamma $; hence the line
$\sigma \ra ({\rm exp}(i\sigma l_e) $ does not cross $W_G^o$ for a generic choice of
the lengths.

\subsection{Asymptotics of $N_G^{\vec{l}}(\ge)$}

 We now prove the first part of Theorem \ref{theo:type2}.

{\it Proof.--}
The proof follows from the same kind of ergodicity argument than in the paper \cite{CdV15}.
More precisely, 
let $w_0 \in W_G^o$ so that $W_G^o$ is of codimension $d(G)$ in
$Y_{\vec{l}}^G $ near $w_0$. 
Thanks to Theorem \ref{theo:tangent} there exists a compact neighborhood $D$ of $w_0$ in $Y_G^{\vec{l}} $
 with smooth boundary, so that $\tau (\ga)$ is  well defined and smooth on $D$.
Then we have, following the argument in \cite{CdV15}, 
\[ N_G^{\vec{l}}(\ge ) \geq {\rm vol} ( \{ \tau \geq -\ge \} \cap D )\]
where the volume is computed w.r. to the Barra-Gaspard measure $\mu_{\vec{l}}=|\iota (\vec{l}) d\ga |$
which is smooth non negative  on $D$. Because $\tau $ vanishes on $W_G^o $ which is of codimension $d(G)$ in $D$, 
this volume is greater than $C\ge^{d(G)/2}$ with $C>0$ (equality holds if the Hessian of $\tau $ at $w_0$ is transversally
 non degenerate).
$ \square $
\section{Bounds for $d(G)$}
\subsection{The  bound  $d(G)\geq 1$}
Let us assume that there is no loops at the vertex $a\in V_0$. 
It follows from \cite{BL16} Theorem 3.6 that for a generic choice of lengths the  eigenvalue $1$  is non degenerate 
and the corresponding eigenfunction does not vanish at $a$. Knowing that $W_G^O$ is semi-algebraic, this implies that 
$d(G)\geq 1$.
\subsection{Upper bounds for $d(G)$}
In the following sections, we will prove the upper bounds for $d(G)$ given in the second part of Theorem
\ref{theo:type2}.
If $\gamma $ is simple cycle of $\Gamma $, we will decompose the vector
$\vec{l}$ as $\vec{l}=(\vec{l_\gamma} ,\vec{ l'})$
where $\vec{l_\gamma } $ is the set of lengths of the edges of the cycle $\gamma $.
Let us recall from \cite{CdV15}, proof of Theorem 4.1. p. 356 (see also Appendix \ref{app:nd})
that, if $\vec{l_\gamma } =(2\pi, \cdots , 2\pi)$, we can always choose $\vec{l'}$ so that $1$ is a non degenerate eigenvalue of
$\Delta ^{\vec{l}}_\Gamma $. The point  $w_0=(1,\cdots, 1, e^{il'_e})$  belongs then to $W_G^o$ and we will study the set $W_G^o$ near
 such a point $w_0$.
A similar study can be done for paths joining two vertices of degree $1$, but we will omit it.

\subsection{The case of circular graphs}

Let us assume that $|\Gamma |$ is a circle. We can always assume that $V_0=V$.
Then $W_G^o $ is a finite set of points of the form $(\pm 1 )$ with the number of $-1$ even.
Hence  $d(G)=\# V_0 -1=g(G)-1$.

From now, we will assume the $|\Gamma |$ is not a circle and $w_0\in W_G^o$ is choosen as described before.

\subsection{The bound   $d(G)\leq g(G)-1$}\label{sec:g(G)}
Let us now discuss the upper bounds for $d(G)$ in terms of the girth (part 2 of 
Theorem \ref{theo:type2}).

Let $\gamma$ be a simple cycle in $\Gamma $ with $g(G)$ vertices  and consider the 
 submanifold (not closed in  general) $W_G^\gamma $ of $W_G^o$ defined by 
$\vec{l_\gamma } =(2\pi, \cdots , 2\pi) $ and $\vec{l'} $ so that $1$ is a non degenerate eigenvalue of
$\Delta ^{\vec{l}}_\Gamma $.
Then $\Delta _\Gamma ^{\vec{l}}$ admits an eigenfunction with eigenvalue $1$ supported on $\gamma$ and vanishing on all vertices 
of $G$. This defines a stratum $W_G^\gamma$ of $W_G^o$ of dimension
$\# E  -g(G)$.
The codimension of $W_G^\gamma$ in $Y_G^{\vec{l}}$ is $g(G)-1$.

Similar estimates hold by using the smallest path  ending at 2 vertices of degree $1$.

\subsection{The bound  $d(G)\leq \# V_0$}\label{sec:v0}

Let us now discuss the upper bounds for $d(G)$ in terms of $\# V_0$ (part 2 of 
Theorem \ref{theo:type2}).
Let $\gamma $ be a simple cycle of $\Gamma $ of length $k$. 
Let us choose
$\vec{l}_0 =( 2\pi, \cdots, 2\pi, \vec{l'})$ (the lengths $2\pi$ occur $k$ times
as lengths of the edges of $\gamma $),
 so that 
$1$ is a non degenerate eigenvalue of $\Delta ^{\vec{l}_0}_\Gamma $ with an eigenfunction which restricts to each edge of 
$\gg$ to $\sin x_e$ and which vanishes outside $\gg$.
For $\vec{l}$ in some neighborhood $U_0$ of $\vec{l}_0$
with $e^{i\vec{l}} \in Z_\Gamma $ we can choose some eigenfunction $u(\vec{l})$
smoothly dependent of $\vec{l}$.
Let us denote by $F$ the map from $\{ w = e^{i\vec{l}} | \vec{l} \in U_0\} \cap  Z_\Gamma$ 
to $\R^{V(\gamma )}$ which associates to $w$ the restriction to $V(\gamma )$ of $u(\vec{l})$, 
and by ${\cal E}$ the subspace of $T_{w_0} Z_\Gamma $ 
of variations $\delta \vec{l}$ of lengths supported by $\gamma$ so that
$\sum_{e\in\gamma }\delta l_e=0$.
The dimension of ${\cal E}$ is clearly $k  -1$.
Let us prove that the differential $L:{\cal E}\ra \R^{V(\gamma )}$  of $F$ 
is injective.
Denoting by $\vec{l}(t)$ a variation of $\vec{l}_0$ and denoting derivatives at $t=0$ by dots,
we get 
\begin{equation} \label{equ:var} \dot{\Delta} u + (\Delta -1) \dot{u}=0~.\end{equation}
We have to prove that $\dot{u} $ cannot vanish on all vertices of $V_0$ unless
$\delta {\vec{l}}$ vanishes. 
Let us assume, by contradiction, that $\dot{u} $ vanishes on $V_0$.
Using Lemma \ref{lemm:der}, and ordering the vertices of $\gg $ using the orientation of $\gg $ as
$V(\gg )=\{ v_1, \cdots, v_k \} $,
we get
$\dot{u}(v_{i+1})=\dot{u}(v_{i})+\delta l_{e_i}$ with $e_i=(v_i,v_{i+1})$.
We get that $\dot{u}$ cannot vanish identically on $V_0$ unless $\delta \vec{l}$ does.
The final result follows from linear algebra:
If ${\cal F}=F'(w_0) ({\cal E})$, $\dim {\cal F} =k -1$ and hence there is at most  one
$a\in V(\gamma )$ so that 
 ${\cal F}\subset \{ v|v(a)=0 \}$.
If $V_0 \subset V(\gamma ) $ with $\# V_0 \leq k  -1$
and $a\notin V_0$,  the map 
$\pi \circ F'(w_0)$ is surjective where
$\pi $ is the canonical projection from
$\R^{V(\gamma )} $ onto  $\R^{V_0} $.

 \subsection{The case $\# V_0 =1$}

Let us say that $a \in V(\Gamma )$ is irregular if,
for all simple cycles $\gamma $  with $a\in V(\gamma )$, the previous space ${\cal F}$ is included in $ \{ v|v(a)=0 \}$.
Using the 2-connectivity of $\gamma $, we see that there is at most one such vertex $a$.
We conjecture that such a vertex never exists. In any case it does not exist if the graph is homogeneous, ie for any
vertex $a$ and $b$, there exists an automorphism of $\Gamma $ sending $a$ onto $b$.

\subsection{The case $V_0 =V$}

If $X\subset E$, we define
\[ {\cal W}_X:=\{ [\vec{l}]\in W_G ~|~ \forall e\in X, ~\exists u\in \ker (\Delta _\Gamma^{\vec{l}} -1), u_{|e} \ne 0   \}~. \] 
Then the family of sets $ {\cal W}_X,~ X \subset E$ is a partition of $ W_G$.
If $[\vec{l}]\in {\cal W}_X$, all lengths of $e\in X$ are multiple of $\pi$.
The eigenvalue $1$ is of multiplicity $b_1(\Gamma_X)$ where $\Gamma _X$ is the subgraph of $\Gamma $
whose edge set is $X$. Moreover, if $b_1(\Gamma _X)=1$, $\Gamma _X$ reduces to that simple cycle.
Hence 
 $ {\cal W}_X \cap W^o_G \ne \emptyset$ if  and only if $X$ is a simple cycle.
The dimension of the corresponding stratum is
$\# E -\# X $ and the maximal dimension of such a stratum is obtained by taking
the cycle of minimal length .
Hence 
 $d(G)=\min  _{\{X ~{\rm simple~cycle}\} } ( \# X) -1  = g(G)-1$.

\section{Examples}

\subsection{Homogeneous 2-connected graphs}
If the graph $\Gamma $ is homogeneous  and 2-connected, there is no exceptional vertex,
hence $d(G)=1$ if $\# V_0=1$.

In the next examples,  all subsets $V_0$  of $V(\Gamma)$ with $\# V_0 < g(\Gamma )-1$
are contained in a simple cycle.
It is known (see \cite{Be73}) that this  follows from the fact that
$\Gamma $ is $(g(\Gamma )-1) $-connected. 

In fact, we can do better using the 
\begin{lemm}\label{lemm:sym}
If $g$ is an automorphism of $\Gamma $ leaving the simple cycle $\gamma $ invariant and with no vertex of $\gamma $ fixed by $g$, then
there is no exceptional vertex on $\gamma $, ie $d(G) \leq \# V_0 $ for all  $V_0 \subset V(\gamma )$.
\end{lemm}
{\it Proof.--} 
We can choose $\vec{l}_0$ invariant by $g$. It follows that the space ${\cal F}\subset \R^{V(\gamma}$
defined in Section \ref{sec:v0}  is invariant by $g$
and cannot be a coordinate hyperplane because $g$ acts without fixed points on $\gamma$.
$\square $

\subsubsection{Complete graphs}

If $\Gamma $ is the complete graph
\[ d(G)\leq \min (2, \# V_0) ~.\]

\subsubsection{Cubes}
All sets of 2 vertices are part vertices of a simple cycle which satisfies the assumptions of Lemma \ref{lemm:sym}, hence
\[ d(G)\leq \min (3, \# V_0 )~.\]

\begin{figure}[hbtp]
\leavevmode \center
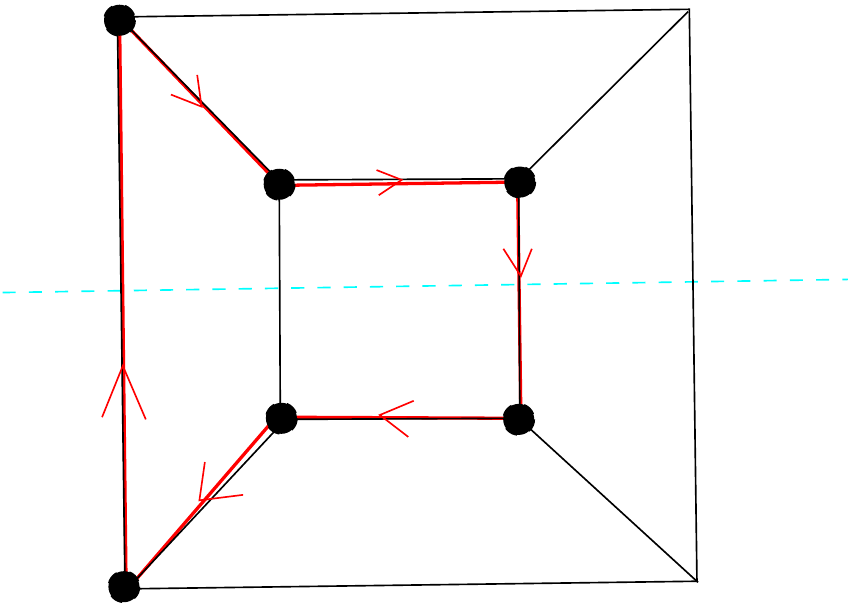
\caption{{\it \small cube with  a simple cycle on $\Gamma$ with a symmetry of order 2 without fixed vertices on $\gamma $.
Any set of 2 vertices of $\Gamma $ is part of such a cycle. }}
 \label{fig:cube}
\end{figure}
\subsubsection{Dodecahedron}
All sets of 3 vertices are part vertices of a simple cycle  which satisfies the assumptions of Lemma \ref{lemm:sym}, hence
\[ d(G)\leq \min (4, \# V_0 )~.\]
\begin{figure}[hbtp]
\leavevmode \center
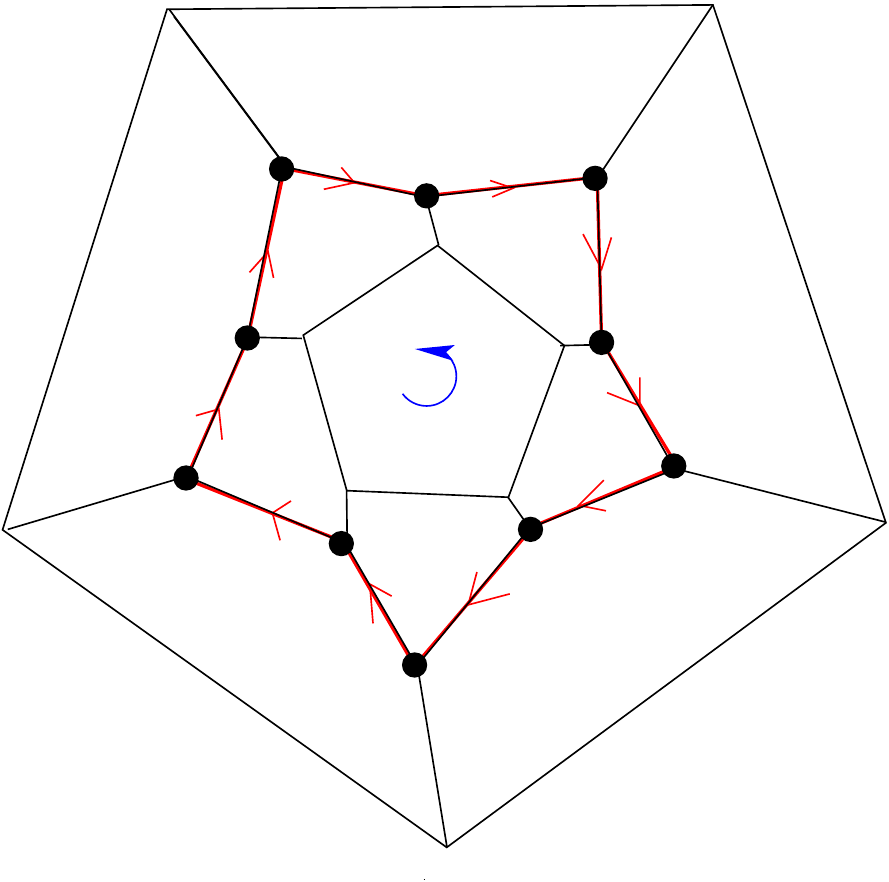
\caption{{\it \small dodecahedron with  a simple cycle on $\Gamma$ with a symmetry of order 5 without fixed vertices on $\gamma $.
 }}
 \label{fig:dode}
\end{figure}
\subsubsection{Petersen graph}

All sets of 3 vertices are part vertices of a simple cycle which satisfies the assumptions of Lemma \ref{lemm:sym}, hence
\[ d(G)\leq \min (4, \# V_0 )~.\]
\begin{figure}[hbtp]
\leavevmode \center
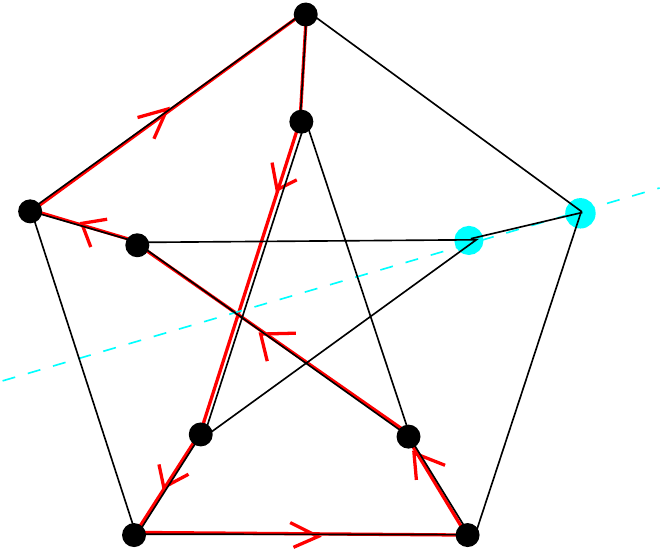
\caption{{\it \small Petersen graph with  a simple cycle on $\Gamma$ with a symmetry of order 2 without fixed vertices on $\gamma $.
Any set of 3 vertices of $\Gamma $ is part of such a cycle. }}
 \label{fig:petersen}
\end{figure}

\subsubsection{Tetrahedron}

The goal of this section is to compute the dimension $d(G)$ in the cases where $\Gamma $ is a tetrahedron.
Instead of making direct computations, we present geometrical arguments which could be extended to other graphs. The link with the 
minor relation between graphs is implicit in our construction of reduced graphs.

We will show the 
\begin{theo}If $\Gamma $ is the tetrahedron, i.e. the clique with four vertices, we have 
$d(G)= \min (\# V_0, 2) $ where $2$ is the girth of $\Gamma $ minus $1$.
\end{theo}
The result follows from Theorem   \ref{theo:type2} for $\# V_0=1 $ and $\# V_0=4 $.
It remains to prove that $d(G)=2$ if $\# V_0=2$ or $3$. The bound $d(G) \leq 2$ comes from Theorem \ref{theo:type2}, 
because the girth of
$\Gamma $ is $3$.

If $[\vec{l}]$ belongs to $W_G^o$ and $u$ is the unique associated eigenfunction
vanishing on $V_0$, let us denote by $n_{\vec{l}} $ the number of edges of $\Gamma $ on which $u$ vanishes identically.
We decompose each stratum of $W_G^o$ following the values of  $n_{\vec{l}} $ into a finite number
of sub-strata. The maximal dimension of these sub-strata
is the same as the dimension of $W_G^o$.
If $[\vec{l}] \in W_G^o$, we have to show that the dimension, denoted ${\rm dim}(\vec{l})$, of the sub-stratum of $W_G^o$ containing 
$[\vec{l}]$ is smaller than $3$.

\begin{figure}[hbtp]
\leavevmode \center
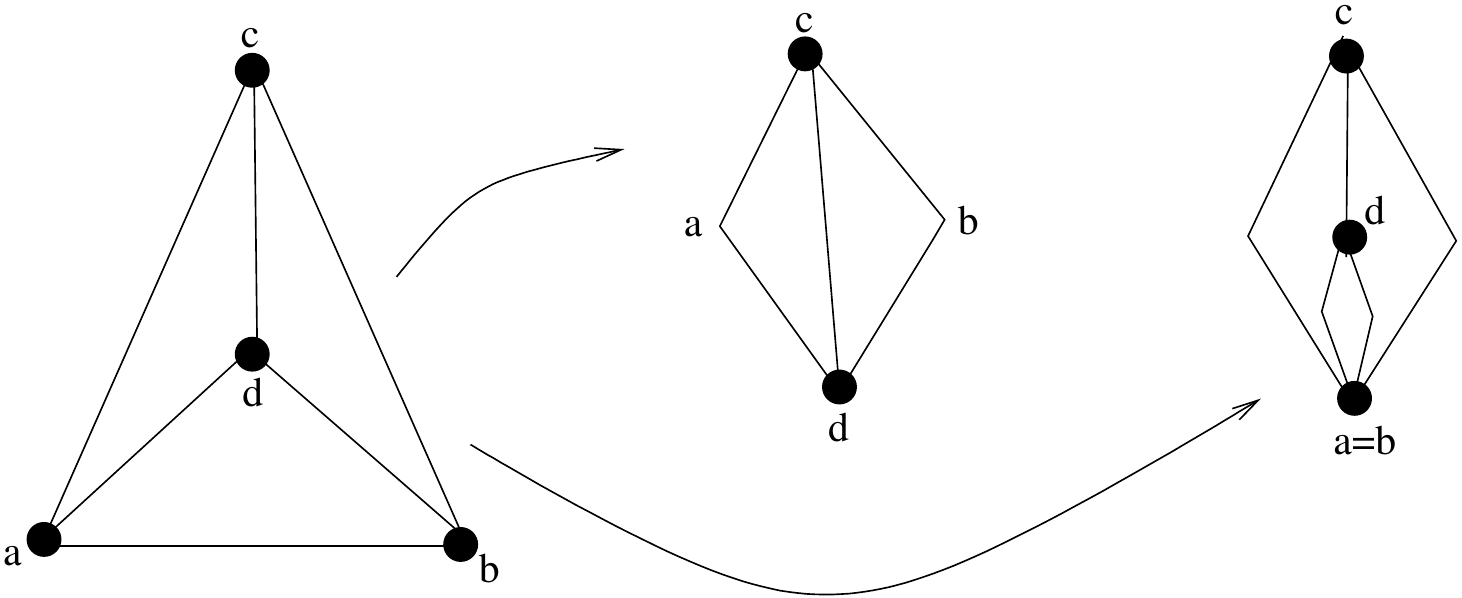
\caption{{\it the two reduced graphs. }}
 \label{fig:reduce}
\end{figure}
Let us discuss the different possibilities:
\begin{itemize} 
\item $n_{\vec{l}}=0 $: 
 \begin{itemize}
\item $\# V_0=2$: let us assume that $V_0= \{ a,b \}$.
The length of the edge $e=\{ a,b \}$ is equal to $0$ modulo
$\pi$. In these cases the function $u$ on $e$ is of the form $ \mu \sin x_e$ with $\mu \ne 0$.
Let us consider the graph $\Gamma ^{\rm red}_c$ obtained by identifying the vertices $a$ and $b$ (contracting the edge $e$).
This reduced graph has $5$ edges.
Denote by $A$ the vertex of degree four obtained by identifying $a$ and $b$.

 \begin{itemize}
\item $l_e=\pi $ modulo $2\pi $:
the Kirchhoff conditions at the vertex $A$ are $u'_1+u'_2+u'_3+u'_4=0$.
We have to look at $W_{G ^{\rm red}_c}^o$ where $V_0=\{ A \}$. The dimension of this manifold is less
than $(5-1)-1=3$: we apply  Theorem \ref{theo:type2} to $G ^{\rm red}_c$
with $V_0=\{ A \}$ and clearly  there is no loop at the vertex $A$. Hence ${\rm dim}(\vec{l})\leq 3$.

\item $l_e=2\pi $ modulo $2\pi $: the Kirchhoff conditions at the vertex $A$ are $u'_1+u'_2=u'_3+u'_4$.
Similar construction gives the same conclusion.
\end{itemize}
\item $\# V_0=3$: in this case  the lengths of the cycle whose vertices are $V_0$ are  equal to $0$ modulo
$\pi$. This implies that the dimension ${\rm dim}(\vec{l})$  of the corresponding sub-stratum   is at most $3$.
\end{itemize}
\item $n_{\vec{l}}=1 $: let us assume that $u$ vanishes identically on the edge $e=\{a,b \}$. 
Let us consider the reduced graph $\Gamma ^{\rm red}_r$ obtained by removing the edge $e$. Topologically, this
 graph has two vertices $c$ and $d$ of degree $3$ and 
and three   edges joigning them.  The dimension of $Z_{\Gamma ^{\rm red}_r}$ is $2$. If $[\vec{l'}]\in Z_{\Gamma  ^{\rm red}_r}^o$,
then the position of the zeroes of the eigenfunction $u$ (unique up to rescaling) on the two edges obtained by removing $a$ and $b$
 determines  all the remaining lengths except
$l_e$. This implies that  ${\rm dim}(\vec{l})=2+1$.

\item If $n_{\vec{l}}\geq 2 $,   one has two cases
\begin{itemize}\item
 If the two  edges  where $u$ vanishes identically have a commun vertex, say $a$,
it follows from the Kirchoff conditions at the vertex $a$ that $u$ vanishes also identically on the third edge with vertex $a$.
Hence the support of $u$ is the remaining cycle (otherwise $u\equiv 0$). This forces the lengths of the edges of this cycle
 to be congruent to $0$
modulo $\pi$ and hence  ${\rm dim}(\vec{l})=3$.
\item If the two edges where $u$ vanishes identically have no commun vertex, then $u$ vanishes at all vertices 
and the four remaining edges have lengths congruent to $0$
modulo $\pi$. This implies that ${\rm dim}(\vec{l})=2$. 
\end{itemize}
\end{itemize}

\subsection{A  simple example where we compute the asymptotics of $N_G^{\vec{l}}(\ge$: the Y-graph}

Let us consider a $Y$ graph with two  edges of length $l$
and $L$ and an infinite edge.
If $z={\rm exp}(ikl),w={\rm exp}(ikL)$, we get
that
\[ {\cal R}_G(z,w)=z^2w^2 -z^2-w^2 -3 ~,\]
whereas $Z_\gG$ is defined by $z=e^{i\ga},w =e^{i\gb }$ with $\ga +\gb =0~{\rm mod~} \pi $.
Since  $W_G= Z_\gG \cap R_G^{\vec{l}}$, with $\vec{l}= (l,L)$ , 
it follows that $W_G=\{  (\pm i, \pm i)  \}$,
  and that
the tangent space to $R_G^{\vec{l}} $ is $d\alpha + d\beta =0$ (due to theorem \ref{theo:tangent}) .
Let us consider a neighboorhood $D$ in  $R_G^{\vec{l}}$ of $w_0 = (i,i)$, namely points $(z={\rm exp}(i{\pi / 2}+i\ga -\tau l),
 w={\rm exp}(i{\pi / 2}+i\gb -\tau L))$ satisfying the resonance equation , with $\ga = +u, \gb = -u $ and $\tau, u$ small .
We get that 
\[  e^{2i(\ga+\gb)} e^ {-\tau(l+ L)} + e^{2i\ga} e^ {-\tau l}  + e^{2i\gb} e^ {-\tau L}=3~,\]
which yields to the following asymptotics
\[ \tau    \sim - {u^2 / 4(l+L)}   ~, \]    as $\tau $ tends to  0 .     
Then the Barra-Gaspard volume $V_\ge$ of the set $D_\ge = D\cap  \{\tau \geq -\ge\}$ is given by
\[ V_\ge = \frac{1}{ 2 \pi^2}\int _{D_\ge}   |L d\gb -ld\ga | = \frac{1}{ 2 \pi^2}\int_{{u^2 / 4(l+L)}\leq \ge}  (l+ L) |du| ~,\]

and thus we  get the following asymptotics
\[ N(\ge) \sim \frac{4 (l+L)^{3/2}}{\pi ^2} \epsilon ^{1/2}~.\]

\subsection{Circular graphs}
A graph $G$ is called circular if $|\Gamma |$ is homeomorphic to a circle.
A general circular graph is denoted 
$C_{N_1,\cdots N_p}$  with $N_i >0 $: this is a subset of $p$ points in circular order  $\{ v_1,\cdots ,v_p \}$
where $N_1$ leads are attached to $v_1$, ...
\subsubsection{$C_1$}
The resonances of $C_1$ are easily shown to be the spectrum of $\Gamma $:
\[ {\rm Res }_{C_1}^L =\{ 2 \pi j /L ~|~j \in \Z \}~.\]

\subsubsection{$C_{1,1}$}

In this case, there are two lengths $l,L$
and putting $z=e^{ikl} $ and $w=e^{ikL }$, 
we get 
\[ {\cal R}_{C_{1,1}}(z,w)= (zw-w-z-3)(zw+z+w-3)~.\]
Then 
 $W_{C_{1,1}}= \{ (-1,-1),~ (1,1) \} $

\subsubsection{$C_{1,1,1}$}
It is still possible to compute
\[ {\cal R}
_{C_{1,1, 1}}(z_1,z_2,z_3)=z_1^2z_2^2z_3^2 -(z_1^2z_2^2+z_2^2z_3^2+z_3^2z_1^2) - 3(z_1^2+z_2^2+z_3^2) -16 z_1z_2z_3 +27 ~,\]
with $z_i=e^{ikl_i }$.
In this case $W_{C_{1,1,1}}$ consists of four points:
$w_1=(1,-1,-1),~w_2=(-1,1,-1),~w_3=(-1,-1,1),w_4=(1,1,1)$.

\subsubsection{$C_{1,\cdots, 1}$}
For all values of $N$, $W_{C_{1,\cdots, 1}}$ is the finite set of
vectors $X$  in $\{-1,+1 \}^N$ so that the number  of negative components of $X$  is even.

\subsubsection{$C_{2}$} This example is studied in \cite{DP11}.
The resonances of $C_2$ are the same as for $C_1$:
\[ {\rm Res }_{C_2}^L =\{ 2 \pi j /L ~|~j \in \Z \}~.\]

\section{Open questions}

The main open question is clearly to get upper bounds of $N_G^{\vec{l}}(\ge)$  as in the conjecture \ref{conj:main}.

Two other questions seem interesting:

\begin{itemize}
\item Is it possible to get an estimate  of the minimal  imaginary part of resonances given by Theorem
\ref{theo:davies} of the 
form $M=M(G)/|L|$ with $M(G)$ depending only of the combinatorics of $G$?
\item Is it true  that if $\vec{l}$ is irrational there is no (or at most a finite number)
of embedded eigenvalues? Compare this to Theorem \ref{theo:emb}.
\item Is the constant $h(G)$ in Theorem \ref{theo:trees} minor monotonic?
\end{itemize}

\appendix
\section{Calculation of a derivative}\label{app:der}

The goal of this section is to prove the
\begin{lemm}\label{lemm:der}Let $\lambda =1$ be a simple eigenvalue of
$\Delta_\Gamma^{\vec{l}_0} $ with an eigenfunction $u_0$.
Let us assume that $u_0$ restricts to some edge $e=[a,b]$  of length $2\pi $ to 
$\sin x$ where $x\in [0, 2\pi ]$ is an arc-length parametrization of $e$ with origine $a$ and end $b$.
If $\vec{l}(t)$ is a smooth deformation of $\vec{l}_0$ so that the eigenvalue $\gl(t)$ is constant equal to $1$, we have
\[ \left(\frac{d}{dt}\right)_{t=0} (u(b)-u(a))= \left(\frac{dl_e}{dt}\right)_{t=0}~.\]
\end{lemm}
{\it Proof.--}
We denote by dots the derivative at $t=0$ and get
\[ (\Delta -1)\dot{u}+ \dot{\Delta }u_0 =0 ~.\]
The restriction of this equation to the edge $e$ gives puting $v=\dot{u}$:
\[ v''(x) + v(x) = \dot{\Delta }\sin x ~.\]
Taking the metric $g(t)= (1+2t)dx^2 $, gives
$\dot{l_e}=2\pi $, while
$\dot{\Delta}= 2 \frac{d^2}{dx^2}$.
The derivative $v$ satisfies
\[  v'' (x)+ v(x)=-2 \sin x ~.\] 
Using the method of variation of constants, we get:
\[ v(x)=\alpha \cos x + \beta \sin x + x \cos x  ~,\]
with some constants $\ga $ and $\gb$, 
and hence 
\[ v(2\pi) -v(0)=2\pi ~.  \]
$\square $

\section{Non degenerated eigenfunctions supported by simple cycles}\label{app:nd}

Let $\gg $ be a simple cycle of $\Gamma $ with ordered vertices
$(x_0, x_1, x_2,\cdots, x_n=x_0)$. Let us consider lengths 
$\vec{l}=( 2\pi, \cdots, 2\pi, \vec{l'})$ where the $2\pi$'s are the lengths of the $n$ edges of $\gamma $
and $\vec{l'}$ the other lengths.
Let us assume in what follows that $\Gamma $ is not homeomorphic to a circle, ie not reduced to $\gamma $.
Let us reprove the following 
\begin{lemm} There exists an open dense subset $\Omega $ of $(\R^+)^{E(\Gamma )\setminus E(\gamma )}$ so that if
$\vec{l'}$ belongs to $\Omega $, then $1$ is a non degenerate eigenvalue of $\Delta _\Gamma ^{\vec{l}}$.
\end{lemm}

The fact that $\Omega $ is open is clear  from general perturbation theory.
Let us start now with $u_0 \in {\cal E}:=\ker (\Delta _\Gamma^{\vec{l_0}}-1)$ the function which restricts to
$\sin t_e$ on each edge of $\gamma $ and vanishes outside.
Let us assume that $\dim {\cal E}>1$ and look at the degenerate perturbation theory of the eigenvalue $1$ while moving
the vector $\vec{l'}$. 
The first order perturbation of the eigenvalues if $\vec{l} $ depends on a parameter $t$ is given by the eigenvalues
of the quadratic form
\[ {\cal Q}:=-\sum _{e\in E} m_e \frac{\pa l_e}{\pa t }_{|t=0}   \]
If the variation of lenths is given by
\[ \vec{l}(t)=(2\pi, \cdots, 2\pi, \vec{l'}_0 + t{\bf 1})~, \]
we get
\[ {\cal Q}=-\sum _{e\in E} m_e \]
which is $<0$  on the orthogonal of $u_0$ in ${\cal E}$.
Hence, for $u$ small, the eigenvalue $1$ is non degenerate.

\section{Semi-algebraic sets}\label{app:alg}
A semi-algebraic  subset of $\R^N$ is a set defined by a finite number of equations and inequations
with polynomial entries.
A fundamental result is the Tarski-Seidenberg Theorem which says
\begin{theo} The image of a semi-algebraic set by a linear map from  $\R^N$ onto $\R^n$
is stil semi-algebraic.
\end{theo}
One of the main properties of semi-algebraic sets is that they admit a stratification: such a set is a union
of a finite number of sub-manifold of $\R^N$, called the strata, so that the closure of each of them is the  union of a finite  number
of strata. The dimension of a semi-algebraic set is then the maximal dimension of these strata.
More details can be found in the classical book \cite{BCR98}.

Let us show as a typical example that $W_G$ is semi-algebraic as well as $W_G^o$.
The equation for eigenfunctions of $\Delta _\Gamma ^{\vec{l}}$ with eigenvalue $1$ is of the
form $M(z)(\vec{a},\vec{b})=0$ with $M(z)$ polynomial in $z={\rm exp}(i\vec{l})$.
The corresponding eigenfunction vanishes at the vertex $x_0$ if $a_e=0$ or $a_e \cos l_e +b_e \sin l_e =0$, depending of the
orientation of  $e=\{x_0,x_1\} $.  
All these equations are polynomial in the coordinates of $z$ and the vectors $\vec{a},\vec{b}$. It is then enough to add $|z_e|^2 =1$.
All of this gives an algebraic sub-set of $\R ^{4\# E }$. The set $W_G$ is the image of this set by the
projection on the $2 \# E$ first factors. 
Concerning $W_G^o$, the uniqueness of the solution of a system of homogeneous linear equations up to
scaling reduces to the non-vanishing of some minors. Algebraicity follows then by taking the sum of the squares of these
minors.

\bibliographystyle{plain}

\end{document}